\documentclass[twocolumn,footinbib,showpacs,preprintnumbers,amsmath,amssymb,pra,superscriptaddress,longbibliography]{revtex4-1}
\usepackage{epsfig,subfigure,amssymb,amscd,amsmath,graphicx,amsfonts,mathtools,mathcomp,pbox}
\usepackage{times,longtable,math dots,hyperref}
\usepackage{soul}

\newcommand{\non}{\nonumber}

\newcommand{\bea}{\begin{eqnarray}}
\newcommand{\eea}{\end{eqnarray}}
\newcommand{\be}{\begin{equation}}
\newcommand{\ee}{\end{equation}}
\newcommand{\ba}{\begin{align}}
\newcommand{\ea}{\end{align}}

\newcommand{\ket}[1]{     |    \,    #1    \rangle}
\newcommand{\bra}[1]{  \langle #1  \,  |} 

\newcommand{\ZZ}{\mathbb{Z}}

\newcommand{\phd}{ {\phantom\dagger} }

\usepackage{tabularx}

\usepackage[usenames]{color}

\DeclareMathOperator{\Tr}{Tr}
\usepackage[usenames,dvipsnames]{xcolor}
\usepackage{soul}
\setulcolor{BrickRed}

\begin{document}

\title{Localization enhanced and degraded topological-order in interacting p-wave wires}

\author{G. Kells}
\affiliation{ Dublin Institute for Advanced  Studies, School of Theoretical Physics, 10 Burlington Rd, Dublin, Ireland}
\author{N. Moran}
\affiliation{ Department of Mathematical Physics, National University of Ireland, Maynooth, Co. Kildare, Ireland}
\author{D. Meidan}
\affiliation{Department of Physics, Ben-Gurion University of the Negev, Beer-Sheva 84105, Israel}
\date{\today}
\begin{abstract}
We numerically study the effect of disorder on the stability of the many body zero mode in a Kitaev chain with local interactions. Our numerical procedure
allows us to resolve the position-space and multi-particle structure of the zero modes, as well as providing estimates for the mean energy splitting between pairs of states of opposite fermion parity, over the full many body spectrum.  We find that the parameter space of a clean system can be divided into regions where interaction induced decay transitions are suppressed (Region I) and where they are not (Region II). In Region I we observe that disorder has an adverse effect on the zero mode, which extends further into the bulk and is accompanied by an increased energy splitting between pairs of states of opposite parity. Conversely Region II sees a more intricate effect of disorder, showing an enhancement of localization at the system's end accompanied by a reduction in the mean pairwise energy splitting. We discuss our results in the context of the Many-Body Localization (MBL). We show that while the mechanism that drives the MBL transition also contributes to the fock-space localization of the many-body zero modes, measures that characterize the degree of MBL do not necessarily correlate with an enhancement of the zero-mode or an improved stability of the topological region.
\end{abstract}

\pacs{74.78.Na  74.20.Rp  03.67.Lx  73.63.Nm}

 \maketitle

The prospects of building quantum devices using topological superconductors has caused a great deal of excitement. In these systems, emergent excitations known as Majorana zero modes that occur at sample edges obey non-Abelian exchange statistics \cite{Read2000,Ivanov2001,Kitaev2001,Kitaev2006,Nayak2008}, and their manipulation is inherently protected from common sources of decoherence. This potentially revolutionary feature has spurred a great deal of theoretical \cite{Fu2008,Lutchyn2010,Oreg2010,Duckheim2011,Chung2011,Choy2011,Kjaergaard2012,Martin2012,NadjPerge2013} and experimental work \cite{Mourik2012,Deng2012,Das2012,Finck2013,Churchill2013,Albrecht2016,Zhang2016,Deng2016,NadjPerge2014,Ruby2015,Pawlak2016}.

The experimental observations in proximity coupled systems  are typically well described within a quasi-particle framework (see e.g \cite{Motrunich2001,Brouwer2011,Brouwer2011b,Akhmerov2011,Rieder2012,Rieder2013,DeGottardi2013,Pientka2013,Stoudenmire2011,Lutchyn2011,Sela2011,Lobos2012,Crepin2014, Hassler2012, Thomale2013, Katsura2015, Gergs2016, Gangadharaiah2011}),  suggesting that at temperatures well below the gap, the properties of these systems are stable to imperfect conditions such as %impurities and 
electron-electron interactions. Recently there have been efforts to understand the stability of these non-abelian excitations  at energies and temperatures well above the topological gap \cite{Gangadharaiah2011,Goldstein2012,Fendley2016,Miao2017,McGinely2017,Kells2015a,Kemp2017,Moran2017,Else2017,Fendley2012,Fendley2014,Jermyn2014,Moran2017,Kells2015b,CommentStrongZeroMode,CommentZn}. These studies directly relate to the effectiveness of symmetry-protected-topological (SPT) systems as platforms for quantum memories. 

In this respect an important recent idea connected to the phenomenon of many-body localization~\cite{Gornyi2005,Basko2006} suggests that the stability of  non-abelian excitations at high energies can be enhanced with additional protection due to disorder-induced localization \cite{Huse2013,Bauer2013,Chandran2014, Kjall2014,Carmele2015,Bahri2015,Wootton2011,Stark2011,Bravyi2012,Potter2016}. 
This notion has been called localization-protected topological-order \cite{LocTop} and its consequences could be far-reaching, allowing for topological quantum processors that can be operated at high temperatures.  Although this would be a remarkable feature, the precise way in which the interplay between disorder and interactions affect the topological order has proved difficult to pin down.  

One complication is that both disorder and interactions are known to be universally detrimental to this symmetry-protected topological-phase. By gradually destroying the superconducting gap which protects it, potential disorder is known to make the Majorana zero modes less localized at the system's boundary. This drives a topological phase transition at a critical strength when the mean free path is half the superconducting coherence length $l_c = \xi/2 $\cite{Motrunich2001,Brouwer2011,Brouwer2011b,Akhmerov2011,Rieder2012,Rieder2013,DeGottardi2013,Pientka2013}.  

Interactions can similarly reduce the topological protection and drive a phase transition to a non-topological phase (see e.g. \cite{Gangadharaiah2011, Stoudenmire2011, Lutchyn2011,Sela2011, Lobos2012, Crepin2014,Gergs2016,Hassler2012, Thomale2013, Katsura2015}). This can be understood in terms of two mechanisms that lift the topological degeneracy associated with the mode: (1) Local charging effects, which give rise to local potentials, can measure the occupation of the zero-mode and (2) Interaction induced decay transitions that change the occupancy of the zero-mode while exciting quasi-particle excitations.  As both disorder and interactions reduce the topological protection, it is reasonable to think that they combine to destroy the topological phase even further. Indeed, analyses of both effects using abelian bosonization suggests that repulsive interactions and disorder do indeed reinforce their destructive effects on the topological  phase  \cite{Lobos2012, Crepin2014}.

These destructive effects add an additional level of complexity to an already difficult numerical problem. This is because in order to demonstrate some enhanced topological protection in the interacting system, one typically needs to obtain precise information about the full many body spectrum, using a statistically significant number of different disorder realisations. Although this full spectrum resolution can be in principle obtained using exact diagonalisation methods, the range of system sizes accessible to this technique is very limited.  This, and the fact that the background negative effects of both interactions and disorder are also strongly present in small systems, makes extrapolation to larger more meaningful systems essentially impossible.

In this manuscript we address these questions from the perspective of many-body-zero-modes \cite{Gangadharaiah2011,Goldstein2012, Fendley2016,McGinely2017,Miao2017,Kells2015a,Kemp2017,Moran2017,Else2017,Fendley2012,Fendley2014,Jermyn2014,Moran2017,Kells2015b, CommentStrongZeroMode,CommentZn} . We focus on the simplest topological superconductor, the Kitaev chain, in the presence of short range interactions and potential disorder.   We study these effects using exact diagonalisation and a numerical procedure that approximates the odd-parity multi-particle steady-states of the interacting commutator $\mathcal{H}=[H,\bullet]$ \cite{Kells2015b}.  In this respect we showcase a key improvement; namely its implementation
using Matrix-Product-Operators (MPO)  \cite{Crosswhite2008} and DMRG-like optimisation \cite{Schollwock2011}.  This super-operator formalism allows us to resolve the position-space and multi-particle structure of the zero-modes as well as to extract statistical information about the entire many-body spectrum. Our analysis shows that, in weakly interacting topological superconductors, disorder can trigger separate effects that both enhance and degrade topological order. As the strength of each mechanism is dependent on the underlying parameter space, this allows for the identification of regimes of parameter space where disorder can degrade (Region I) or improve (Region II) the underlying topological protection of the zero mode.

The structure of  the manuscript is as follows. In section~\ref{sect:sapp} we
review the Kitaev chain (or p-wave wire) model and qualify our central results using both band-projection and perturbation theory. In this section we also review the key results pertaining to MBL and their connection to so called many body zero modes. In section \ref{sect:numerical_methods} we discuss our MPO numerical methodology and examine the connection between the structure of the zero-mode expansion and the statistical estimates of the pairwise energy level splitting. In section \ref{sect:numerical_results}  we outline the numerical results themselves. 

We also include several appendices: In Appendix A we discuss the perturbative case for zero-modes.  In Appendix B we discuss our MPO algorithm and add more details to the error analysis provided in the main text.  In Appendix C we provide the results from exact diagonalization calculations. In Appendix D we outline the formal construction of many-body zero modes and discuss the relationship between energy relaxation processes and resulting multi-particle structure of the zero-mode position space expansion.

\section{Model and  physical picture}
\label{sect:sapp}

We formulate our results using the lattice p-wave superconducting  model or Kitaev chain \cite{Kitaev2001}:
\begin{align}\label{H:Kitaev}
H_0&=&-\sum_{j=1}^N \mu_j (c_{j}^\dag c_ {j} -\frac{1}{2}) - \sum_{j=1}^{N-1} t c_{j}^\dag c_{j+1} 
 + \Delta c_{j}^\dag c_{j+1}^\dag +{\rm h.c.} 
\end{align}
where coefficients $t$, $\Delta$ and $\mu_i$ are for the hopping, pairing
amplitude and local chemical potential at site $i$ respectively. To model
disorder we allow the local chemical potential to vary around an average value
$\mu$ with the standard deviation set by the parameter  $\lambda$. The normal-state mean-free-path is given as
$l= \frac{v_F^2}{\lambda^2}$, where $\hbar v_F = 2ta \sqrt{\frac{\mu+2t}{t}}$ is the Fermi-velocity. 

Interactions are included through the local quartic term
\begin{equation}
\label{H:Int}
H_{I} =2U \sum_{j=1}^{N-1} \left(c_j^\dag c_j -\frac{1}{2}\right) \left(c_{j+1}^\dag c_{j+1} -\frac{1}{2}\right). 
\end{equation}
The phase of the p-wave superconducting pairing potential can be chosen to be
real. When $|\Delta|>0$ and $|\mu| < 2t$ the $H_0$ system is known to be in a
topological phase with Majorana zero modes exponentially localised at each end
of the wire\cite{Kitaev2001}. In what follows it is useful to work in a
basis of position space Majorana operators defined as:
\begin{equation}
\label{eq:Majdef1}
\gamma_{2j-1} =i (c^\dagger_j - c_j^\phd ), \gamma_{2j} = (c^\dagger_j + c_j^\phd ).
\end{equation}
These obey $\{\gamma_i ,\gamma_j \}= 2 \delta_{ij}$ and thus $\gamma_i = \gamma_i^\dagger$ and $\gamma_i^2 =I$.  

\begin{figure}[htb]
\centering
\includegraphics[width=0.5\textwidth,height=0.45\textwidth]{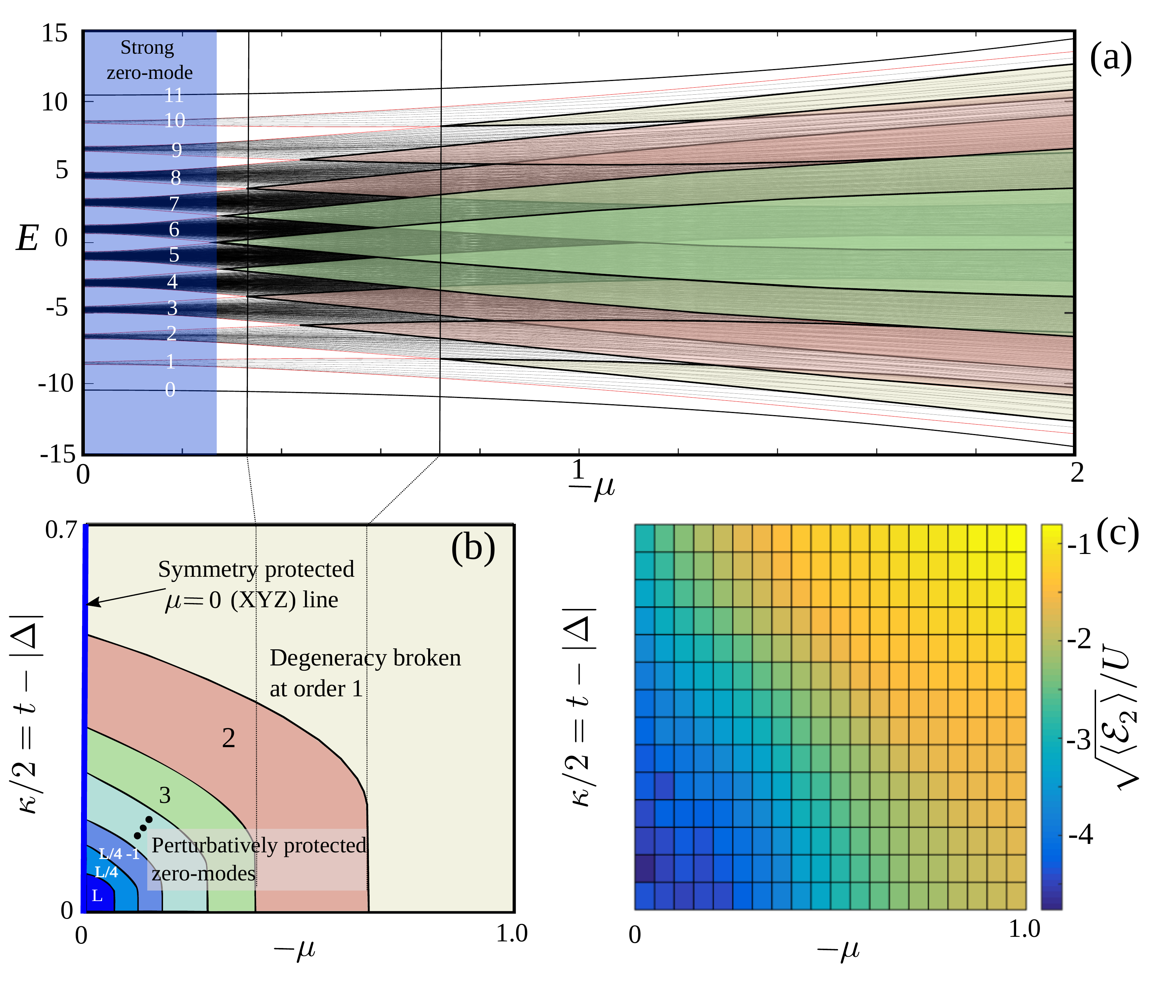}
\caption{ Figure (a) : the full many body spectrum of a small $L=12$ system with $\Delta=0.9 t$ and  $U=0$ . We label bands according to  the number of bulk (non-zero mode) excitations above the ground state.  When band crossing occur, interactions can induce avoided level crossings between bands with different fermion number and different occupation of the zero mode. These are generally different in even and odd sectors; causing the degeneracy associated with the zero mode to lift. Away from the flat band limit $\mu=0$ and $ t=|\Delta|$, bands in the middle of the spectrum begin to cross. Interactions induced transitions require the excitation of a large number of quasi-particle excitations in this limit. Figure (b): phase diagram constructed by outlining where crossings first occur for a $L=50$ system.  It also indicates a key exception along the $\mu=0$ line, where the crossings are protected by additional symmetry \cite{Fendley2016}.   Figure (c): $ \sqrt{\langle \mathcal{E}_2} \rangle/U$ given by   \eqref{eq:e2} in the clean system with $U=0.1t$, calculated using the MPS/DMRG algorithm for a system of length $L=50$.   Unless specified otherwise, all energy scales are in units of $t$.} 
 \label{fig:zeromodes}
 \end{figure}

%Our central results are qualified using both band-projection and perturbation theory. 
When interactions are absent, the many body spectrum is doubly degenerate. The two states that form an almost degenerate pair differ by the occupation of the zero mode, made up of two Majorana  bound states  exponentially localized at the two ends of the chain. The energy splitting  between pairs  depends on the spatial decay rate of the Majorana zero modes and is given as $\delta \sim e^{-L/\xi}$ where $\xi \sim t/\Delta$ is the superconducting coherence length.  

Interactions can lift the two fold degeneracy in two ways. Firstly, by introducing local charging effects, which can measure the occupation of the zero mode.  As information of the occupancy of the zero mode is stored non-locally, this lifting  occurs  at an order of the interaction strength $U $ which scales with the system size $\sim U^L $.  Crucially, interactions can also change the occupancy of the zero mode by introducing energy relaxation processes whereby  a finite energy excitation can decay into the zero mode while exciting a pair of quasi-particles. These decay processes serve as a lifetime for non-interacting states, which can be estimated from a Fermi golden rule type analysis. The simplest lowest-order decay process  is the transition of  a  quasi-particle excitation to two quasi-particle excitations, while changing the occupancy of the zero mode (leaving all other quasi-particle excitations unaltered):
\be
\Gamma \sim \frac{|U|^2}{\Delta \epsilon}
\ee
where $ \Delta \epsilon= 2\epsilon^{0/1}_{min}-\epsilon^{1/0}_{max}$ where the superscript denote the state of the zero-mode, $2\epsilon^{0/1}_{min}$ is the minimal energy of a state with two bulk quasi-particle excitations and $\epsilon^{1/0}_{max}$ is the maximal energy of a state with a single bulk quasi-particle excitation.  

Our main insight is then based on the fact that in a clean system there are regions of parameter space where these real decay transitions are  suppressed, we refer to this regime, as Region I.  In the clean non-interacting limit, Region I can be defined by the requirement that  $\Gamma <  \Delta \epsilon $ which can be written as:  [see Appendix \ref{sect:perturbzero} for a detailed discussion]:
%\bea
%\label{eq:R1def}
% && -\mu  < \frac{2}{3} t, \quad \quad \quad \quad \quad  \quad \text{if } \frac{t}{2} (-\mu+2t) < |\Delta|^2\\
% && -\mu  < 2 ( |\Delta| \sqrt{4- \frac{\mu^2} {t^2 -|\Delta|^2}} -t) ,  \text{otherwise}  \non
%\eea
\bea
\label{eq:R1def}
 && |\mu|  < \frac{2t-U}{3} , \quad \quad \quad \quad \quad  \quad \text{if } |{\mu} |> \frac{2}{t}(t^2-\Delta^2)\\
 && |\mu|  < 2  |\Delta| \sqrt{4- \frac{\mu^2} {t^2 -\Delta^2}} -2t-U ,  \text{otherwise}  \non
\eea
The complement space, where  $\Gamma >  \Delta \epsilon $, is identified as Region II.  We remark that while Region I can be prominent in lattice models,  experimental  realizations of the Kitaev chain are typically characterized by  weak proximity coupling $ \Delta\ll t$, and the parameter space is dominated by Region II.

Disorder modifies this picture in three ways by:  (1) increasing the coherence length $\xi$, making the Majorana zero modes less localized at the systems boundary  \cite{Motrunich2001,Brouwer2011,Brouwer2011b,Akhmerov2011,Rieder2012,Rieder2013,DeGottardi2013,Pientka2013}, (2) broadening the width of the single particle excitation band and (3) decreasing the localisation length of bulk excitations \cite{Anderson58}.

\begin{figure}
\includegraphics[width=.4\textwidth,height=0.3\textwidth]{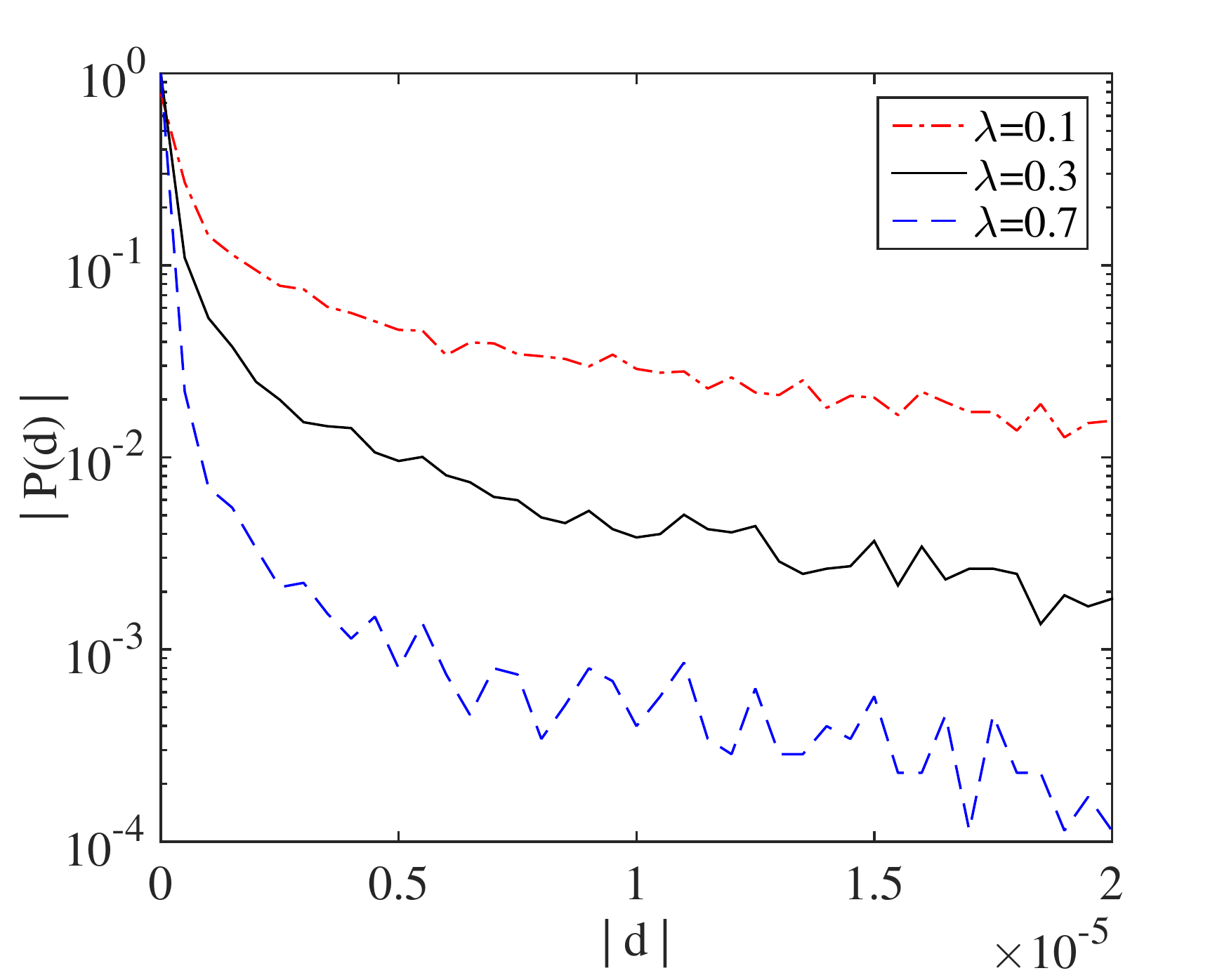}
\caption[]{ The probability distribution of the first-order interaction-induced decay-amplitude  $|d| = |\bra{n} H_I \ket{m}|$ decays as we increase the disorder strength $\lambda$.  Here $\ket{m}$  is a non interacting state with two bulk excitations and the zero mode unoccupied and  $\ket{n}$ is a non interacting state with a single bulk excitation and the zero mode occupied.  The plots were obtained in Region II for $\mu=-t$, $\Delta=0.5 t$, $U=0.1 t$ and for a wire of length $L=100$ . Increasing the value of the disorder parameter $\lambda$ shifts the distribution towards zero splitting.  }
\label{fig:distrib1}
\end{figure}

On a single particle level, both Regions I and II experience  a similar effect of disorder which extends the zero mode operator further into the bulk, thus gradually lifting the degeneracy that protects the topological phase.   However, on top of this single particle effect, disorder plays a much more subtle role:  In Region I disorder has a universally adverse effect because, by also broadening the bulk single particle excitation band, it also drives the system towards a regime where  decay transitions can occur.  Although disorder may also increase the number of  decay transition in Region II,  in this case the energy splitting associated with these decay transitions is reduced, as shown in Figure \ref{fig:distrib1}. This behaviour is directly connected to the spatial localization of the bulk states, which suppress these decay processes within a localization length  (see e.g. \cite{Gornyi2005,Basko2006,Huse2013}). 

These competing behaviors are revealed in the numerical analysis  (see section \ref{sect:numerical_results}) of the multi-particle structure where we see clear evidence of both localization-enhanced {\em and} localization-diminished topological-order. Near the system edges, in both Regions I and II,  disorder increases the decay-length of single particle terms as well as locally clustered components. However,  further from the sample edge, the spatial decay of the locally clustered components shows a clear distinction between the two regions of phase space, as highlighted in Fig. \ref{fig:u3plotsl0} (c) and (d). In Region I all local clusters extend further into the bulk in the presence of disorder. Conversely,  Region II exhibits a transition from non-decaying local-clusters (see \ref{eq:Majspl}) in the clean system to exponentially decaying in a disordered medium. 

We will also show how the aforementioned decay of local multi-particle clusters is reflected in the  mean energy-level-splitting $\langle \delta \rangle$ between pairs of states from opposite parity sectors. Region I, which is dominated by the single particle behavior, exhibits predominantly localization-diminished topological order. Conversely, Region II displays localization-enhanced topological order at moderate disorder strength. 

\vspace{2mm}
{\it Relation to many body localization}
The observed enhancement of topological protection is related to, but distinct from, Many-Body-Localization (MBL) \cite{Gornyi2005,Basko2006}. In particular, while our  analysis shows that disorder induces opposing effects on the zero mode structure in the two regimes of parameters,  measures that characterise the extent of MBL show no prominent differences between the two regions. Nor do they demonstrate any noticeable change as the topological order is destroyed. 

The transition to an MBL phase  can be understood as a dynamical phase transition, in the sense that one can construct an extensive number of integrals of motion \cite{Huse2014,Chandran2015a,Serbyn2013a} that constrain the dynamics of the system to a degree that it does not thermalize in the way expected from the eigenstate thermalisation hypothesis (ETH) ~\cite{Deutsch1991,Srednicki1994,Srednicki1996}. These features of the  ETH-MBL transition give rise to a rich variety of signatures, which can be detected in the level statistics \cite{Bauer2013,Oganesyan2007,Pal2010,Cuevas2012,Laumann2014,Luitz2015}, in the entanglement entropy \cite{Bauer2013,Kjall2014,Grover2014}, in the response \cite{Berkelbach2010,BarLev2015} and  in the dynamics of the system under consideration \cite{Znidaric2008,Bardarson2012,Serbyn2013}. 
 
To address the possible association between enhanced topological order and the ETH-MBL transition we  focus on one sensitive characterization of the transition that is based on the eigenvalues of the generalised single-particle density matrix 
\be
R_n=\left[ \begin{array}{cc}   \rho & \kappa  \\ \kappa^\dagger & I-\rho  \end{array} \right] ,
\ee
where $\rho_{ij} = \bra{n} c_i^\dagger c_j \ket{n}$, $\kappa_{ij} = \bra{n} c_i^\dagger c_j^\dagger \ket{n}$, and $\ket{n}$ is a many body eigenstate.  Similarly to Ref. \onlinecite{Bera2015}, for which the system was not superconducting and so $\rho $ was sufficient, the eigenvalues of $R_n$  constitute the occupation spectrum and this exhibits distinct  behaviour in the two phases. In the delocalized (ETH) phase, they are  expected to be close to the mean filling fraction, while in the localized phase (MBL) they should tend to their asymptotic values $\in\{0,1\}$. Consequently, it is possible to characterize  the transition to an MBL phase by a step-like jump in the occupation spectrum.  

\begin{figure}
\includegraphics[width=.4\textwidth,height=0.3\textwidth]{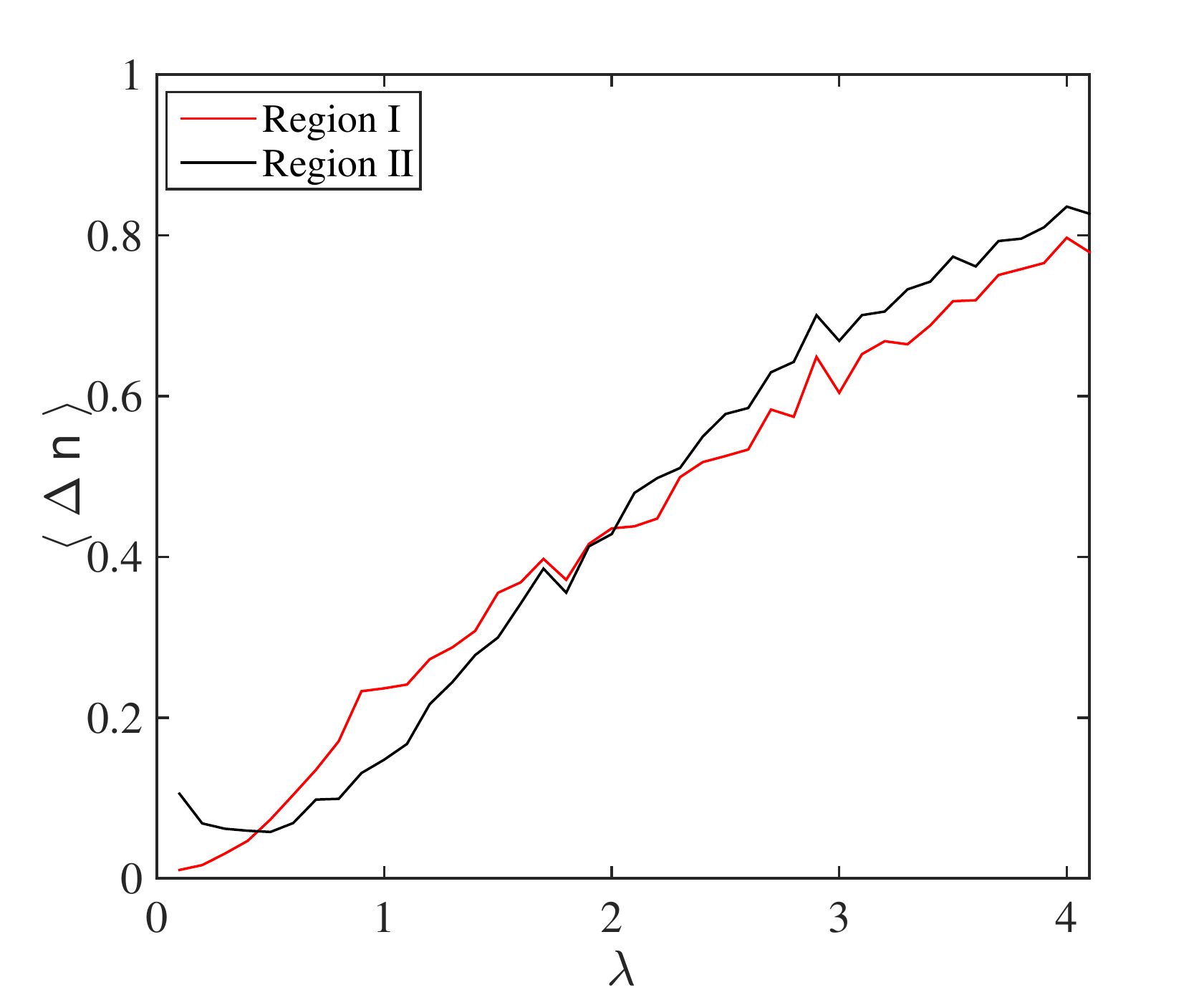}
\caption[]{ The dependence of the discontinuous jump of the  occupations of natural orbitals,  with increasing disorder strength $\lambda $ for Region I ($\Delta = 0.7 t $, $ \mu=-0.2 t$ and $U=0.1t $, red curve) and Region II ($\Delta = 0.5 t $, $ \mu=- t$ and $U=0.1t $, black curve).
Following Ref. \onlinecite{Bera2015}, in the delocalized phase, the occupations are expected to be close to the mean filling fraction, while in the localized phase they tend to their asymptotic values $\langle n_N\rangle =\{0,1\}$, and the occupation spectrum exhibits a discontinuous jump $\Delta n = n_{L+1}-n_{L}$. Consequently, the averaged value of the discontinuous jump in the occupation  spectrum can be used to characterize the  ETH-MBL transition. While disorder is shown to induce opposing effects on the zero mode structure in the two regime of parameters, see Fig. \ref{fig:u3plotsl0},  its effect on the  occupation spectrum is essentially identical.  The data shown is for a system of length $L=15$ and each data point in averaged over 100 disorder realisations with a sampling of 100 states per realisation (50 in each sector) around $E=0$. 
}
\label{fig:MBL}
\end{figure}

In Fig. \ref{fig:MBL} we show the value of the discontinuous jump in the occupation spectrum of the single particle density matrix, in Region I (red curve) and Region II (black curve) for a system of size $L=15$.  Although disorder induces opposing effects on the zero mode structure in the two regimes of parameters,  this MBL measure does not distinguish between the two regions.  Moreover it is also insensitive to the underlying topological order which, for the representative parameters for Regions I and II, is destroyed by disorder strength $\lambda \gtrapprox 2.9$ and $2.3$ respectively. 

The distinction between localization and the observed enhancement of topological protection is twofold. Firstly, while localization in Fock space is known to suppress decay transitions, not all transitions are detrimental to the zero mode. Consequently, localisation induced protection can only occur in regions of phase space where these harmful decay process are abundant. This corresponds to our definition of Region II. As standard measures of localization cannot distinguish transitions that couple states with different occupation of the zero mode and those who do not, these cannot pick up the difference between the two regimes of parameter, as we show in Fig  \ref{fig:MBL}.  Secondly, it is not clear that the topological superconducting phase survives strong potential disorder. That is to say, in the limit when the system breaks down into segments of localization length, topological protection can be lifted altogether due the small size of each segment as compared to the superconducting coherence length, which is known to increase in the presence of potential disorder. This single particle effect is crucial to the suppression of topological protection in topological superconductors, but plays no role in the localization transition. It is for this reason that calculations aimed at detecting the MBL transition (such as the one showed in Fig  \ref{fig:MBL}), are insensitive to the disorder induced topological phase transition.
 
\section{Numerical methods}
\label{sect:numerical_methods}
We seek  to identify an operator  which:
\begin{enumerate}
\item Commutes with the Hamiltonian, up to small corrections:
$ [H,\Gamma]\sim 0$.
\item Anticommutes with the total parity: $ \{P,\Gamma\}=0$.
\item Is Hermitian: $ \Gamma= \Gamma^\dag$.
\item Is its own inverse: $\Gamma^2=I$.
\end{enumerate}

Our numerical procedure is based on giving matrix representations to the commutator $\mathcal{H} =[H,\bullet]$ using the operator (Hilbert-Schmidt) inner product \cite{Goldstein2012,Kells2015b}. This procedure  is based on what is called the Choi-Jamiolkowski-isomorphism \cite{Choi1972,Choi1975,Jamiolkowski1972}, also referred to as Third Quantization \cite{Prosen2008}. The numerical algorithm itself can be seen as a hybridization of the variational position-space algorithm applied in Ref. \onlinecite{Kells2015b}, and methods that represent super-operators such as $\mathcal{H}$ (or more generally the Limblad super-operator) as Matrix-Product-Operators (see e.g. \cite{Mascarenhas2015,Cui2015} )  

In the presence of interactions, this procedure produces a many body operator of the general form:
\begin{equation}
\label{eq:Majspl}
\Gamma= \sum_i^{2L} u^{(1)}(i) \gamma_i + \sum_{ijk}^{2L} u^{(3)}(i,j,k) \gamma_{i} \gamma_{j} \gamma_{k} + ... 
%\gamma^R(U) = \sum_i^{2L} u^{(1)}_R(i) \gamma_i + \sum_{ijk}^{2L} u^{(3)}_R(i,j,k) \gamma_{i}  \gamma_{j} \gamma_{k} + ... \non
\end{equation}
where $\gamma_i $ is a Majorana operator at position $ i$ and $u^{(n)}(i,j,k,..)  $ is the coefficient of the $n$ particle term, with $n$ majorana modes located at positions $i<j<k$.  We numerically calculate the zero mode wave function by identifying the operators that minimize the expression $\Tr( \Gamma_{L/R}^\dagger \times[ H, [H, \Gamma_{L/R}]] )/2^L$, subject to constraints 2-3, and where $ \Gamma_{L/R}$ stands for zero mode localized at the
left/right side of the chain. Although constraint 4 is not actively enforced, our methodology insures that it is approximately obeyed.

{\em Statistics of level splittings and error estimates: } 
In addition to probing the local structure of the zero mode  \eqref{eq:Majspl}, the methodology described above allows us to estimate the average level splitting between pairs of states from different parity sectors. To see this we first examine the Hamiltonian in the system eigenbasis. In the topological phase the many body spectrum is doubly degenerate up to corrections $\delta_n$:
\begin{align}\label{H:many_body}
\nonumber
H&= \sum_{n} E_n \left[|n_1\rangle \langle n_1|+|n_0\rangle \langle n_0|\right]\\
&+\delta_n/2 \left[|n_1\rangle \langle n_1|-|n_0\rangle \langle n_0|\right].
\end{align}
Here $0/1$ refers to the occupation of the zero-mode. In the non-interacting system, the zero-mode operators
are eigenmodes of the Hamiltonian, $\left[ |n_0\rangle \langle n_1|,H\right]= \delta \left(|n_0\rangle \langle n_1|\right)$, which means that the {\it many body} spectrum consists of pairs of states distinguished by the occupation of the zero mode, and displaced by a uniform energy splitting: $\delta_n = \delta \sim e^{-L/\xi}$. In general, however, interactions give rise to a distribution of pair splitting $P(\delta_n )$ which are not necessarily exponentially small. 

In this basis the MPS/MPO methodology constructs an {\it approximate} Majorana zero mode with the following structure
\begin{align}\label{eq:MBZM1}
\Gamma_{L} &=  \sum_{n} (1-\alpha^{L}_n) \left[ |n_0\rangle \langle n_1| + |n_1\rangle \langle n_0| \right]\\ 
\non &+  \sum_{n \ne m} \beta^L_{nm}  \left[ |n_0\rangle \langle m_1|+ |m_1\rangle \langle n_0| \right]\\
\label{eq:MBZM2}
\Gamma_{R} &=-i \sum_{n} (1-\alpha^{R}_n) \left[ |n_0\rangle \langle n_1| - |n_1\rangle \langle n_0| \right]\\ 
\non  &-i \sum_{n \ne m} \beta^R_{nm}  \left[ |n_0\rangle \langle m_1| - |m_1\rangle \langle n_0| \right]
\end{align}
where  $\alpha$- and $\beta$-terms represent diagonal/off-diagonal errors respectively. The commutator of the near zero mode operators $\Gamma_{L/R} $ with the interacting Hamiltonian, allows to estimates the energy level statistics:
\begin{align}
\label{eq:e1}
\mathcal{E}_1 &=i \Tr( \Gamma_L \times [H, \Gamma_R] )/2^L =  \langle \delta \rangle + \chi_1 \\
\label{eq:e2}
 \mathcal{E}_2&=\Tr( \Gamma_L\times[ H, [H, \Gamma_L]] )/2^L = \langle \delta^2 \rangle   +\chi_2
\end{align}
where 
\begin{align} 
\label{eq:chi1}
\chi_1&= - \frac{1}{2^L}  \sum_n \delta_n \left[\alpha^{L}_n +\alpha^R_n  - \alpha^{L}_n \alpha^{R}_n \right] \\ 
\non &+  \frac{1}{2^L} \sum_{n\ne m} \beta^L_{nm} \beta^R_{nm} [E_n-E_m+\frac{\delta_n}{2}+\frac{\delta_m}{2}] 
\end{align}
and
\begin{align}
\label{eq:chi2}
\chi_2&= - \frac{1}{2^L}  \sum_n  \delta_n^2 ( 2 \alpha^L_n- (\alpha^L_n)^{2}) \\ 
\non&+  \frac{1}{2^L} \sum_{n\ne m}( \beta^L_{nm})^{2} [E_n-E_m+\frac{\delta_n}{2}+\frac{\delta_m}{2}]^2. 
\end{align}%  in Appendix \ref{sect:dmrgsuper}. 
Crucially we note that as Eq. \eqref{eq:e1}  involves the two near zero modes that are predominantly supported on opposite ends of the system, its value is influenced by the degree of localization of local-clusters of the constituent Majorana components.  This is unlike Eq. \eqref{eq:e2}  for which we only need either $\Gamma_L$ or $\Gamma_R$.   In addition, as the expression for the error $\chi_1$ is an average over contributions of random sign, we expect it to be small.  In contrast the error in the $\mathcal{E}_2$ estimate, which is what the DMRG routine is trying to minimise, consists of  positive definite contributions which do not  cancel. These suggest that   $\chi_2 $ may be substantial and possibly dominate the estimates for $\mathcal{E}_2$. Evidence supporting this conjecture is provided in Appendix \ref{sect:dmrgsuper}.

The search for a zero-mode operator is similar in some ways to the search for integrals of motion (IOM) which have a finite position space support in the MBL regime, 
see for example \cite{Kim2014,Ros2015,Imbrie2016,You2016,Geraedts2016,OBrien2016,Ilievski2016,Friesdorf2015,He2016,Rademaker2016a,Rademaker2016b,Pekker2017,Quito2016,Khemani2016,Chandran2015b,Monthus2016}. Nonetheless there are some important distinctions. In parity preserving systems such as the one under study, operators like $\ket{n}\bra{n}$ (or any superposition thereof) have multinomial expansions that are sums of even terms only. In contrast, in the search for a zero mode  we are essentially looking for two IOMs that switch the parity of state being acted on, see for example Eq. \eqref{eq:MBZM1}. In the representation of our MPO encoding this is  enforced by constraining the multinomial expansion to contain odd numbers of fermion terms only. As a result of this, these two fermionic IOMs anti-commute ($\{\gamma_L,\gamma_R\}=0$) with each other and with the Parity operator  ($\{\gamma_{L/R},P\}=0$).

The odd excitation sector of the super-operator $[H,\bullet]$ differs from the even excitation sector in that it only contains an IOM when there is an underlying degeneracy (in the Hamiltonian) between a state in the even sector and a state in the odd sector Although this may happen by accident between any pair of states, a strong many-body mode would ensure it approximately happens between $ 2^{L-1}$ pairs simultaneously. In this respect the constraints that $\Gamma^2=I$ ensures that there is approximately equal weight given to each diagonal outer product $ \left[ |n_0\rangle \langle n_1| \pm |n_1\rangle \langle n_0| \right]$ in summations Eqs. \eqref{eq:MBZM1} and \eqref{eq:MBZM2}.

\section{Numerical Results}
\label{sect:numerical_results}
{\it Decay rates - Clean case:}
We now begin our discussion of numerical results, starting with single and multi-particle decay rates of near zero-modes. 
Looking at Eq. \eqref{eq:Majspl}, in the non-interacting system, the multi-particle expansion coefficients $u^{(n)}=0$ for all $n>1$ and  the spatial profile of $ u^{(1)} (x)$ decays exponentially with $t/\Delta$.   In order to generalize this spatial profile for the multi-particle components of the many body zero mode, $u^{(n)}(i,j,k,..)   $ which depend on multiple position indices, we have calculated the spatial profile of two representations of the three particle component corresponding to local $u^{(3)}_{\rm l}(x)=u^{(3)}(\gamma_{2x-1},\gamma_{2x+2},\gamma_{2x+3})$ and non-local  clusters $u^{(3)}_{\rm nl}(x) = u^{(3)}(\gamma_1,\gamma_{2 x},\gamma_{2x+1})$ \cite{clusters}. 

\begin{figure}[!htb]
\centering
\includegraphics[width=0.48\textwidth,height=0.37\textwidth]{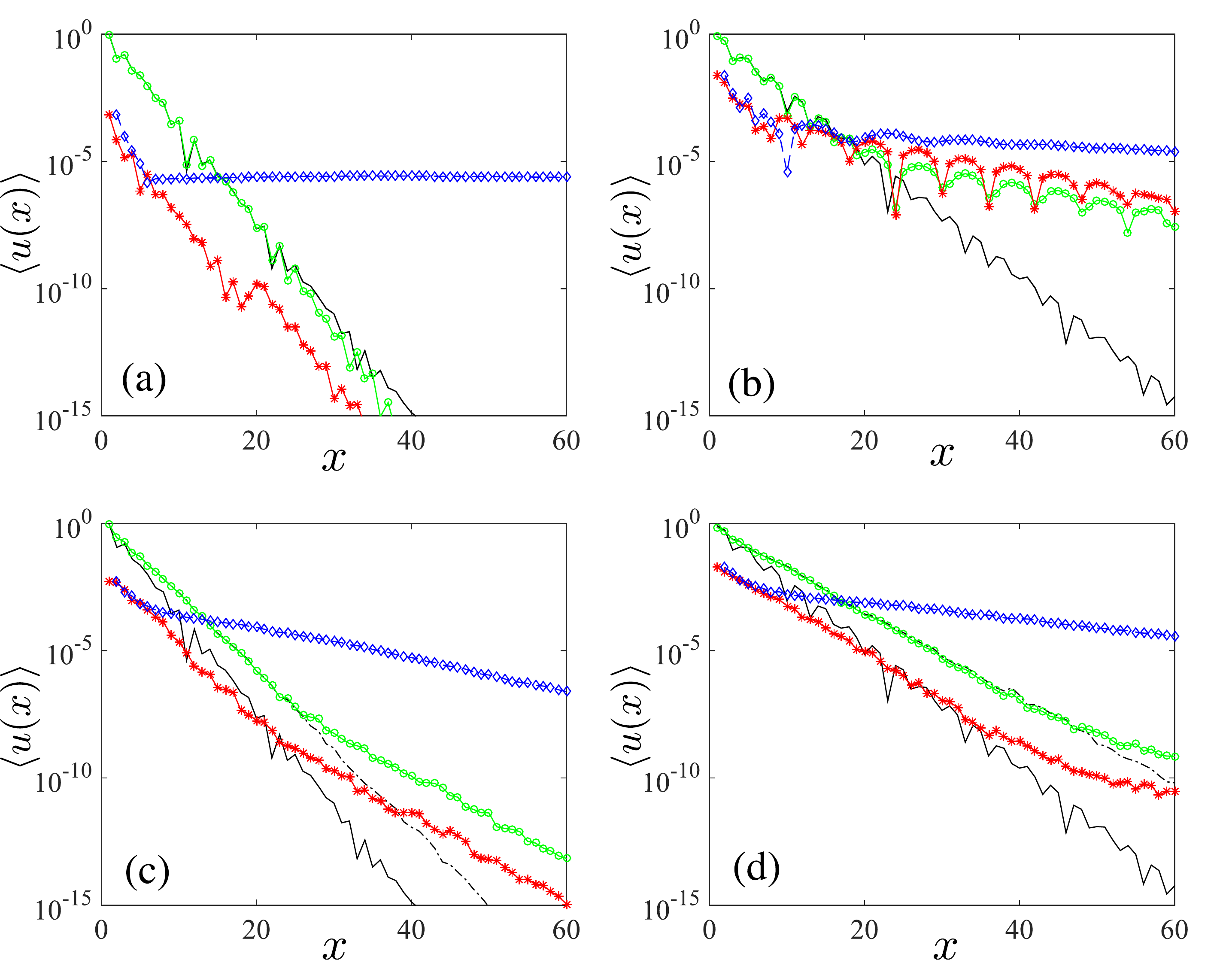}
\caption{Spatial composition of the single particle component $u^{(1)}(x)$ (green circles) and the three particle components composed of   local $u^{(3)}_{\rm l}(x)=u^{(3)}(\gamma_{2x-1},\gamma_{2x+2},\gamma_{2x+3})$ (red stars) and  non-local  clusters $u^{(3)}_{\rm nl}(x) = u^{(3)}(\gamma_1,\gamma_{2 x},\gamma_{2x+1})$  (blue diamonds) clusters for an $L=100 $ site chain. Figures (a) and (c)  where calculated for Region I  ($\Delta=0.7t $ and $ \mu=-0.2t$), in the clean ($ \lambda=0$) and disordered ($ \lambda=0.7$) case, respectively. Figures (b) and (d)  were calculated for Region II ($\Delta=0.5t $ and $ \mu=-t$), in the clean ($ \lambda=0$) and disordered ($ \lambda=0.7$) case, respectively. All plots where calculated for $ U=0.1 t$ with a bond dimension of $\chi=128$.  In  Region I (Fig. (a)) local clusters $u^{(3)}_{\rm l}$  decay with the same rate as $u^{(1)}$ which itself resembles the non-interacting wave function (solid line). In Region II (Fig. (b)) $u^{(3)}_{\rm l}$ decays much more slowly than $u^{(1)}$ and seems to follow a power law. Moreover, as  one moves away from the edge, the single particle decay rate starts to follows that of the local three particle clusters $u^{(3)}_{\rm l}$. Disorder has strikingly different effects in the two Regions. In  Region I, (Fig (c)) disorder extends both the single particle components $u^{(1)}$  and the local clusters of Majorana $u^{(3)}_{\rm l}$ which both initially  follow the non-interacting disorderd decay (dashed dotted line) and eventually follow a slower decay length. Conversely, in Region II, (Fig. (d)) disorder substantially suppresses the spatial extent of locally clustered components. }\label{fig:u3plotsl0}
 \end{figure}

In a clean system we observe notable distinctions between Regions I and II , see Figure \ref{fig:u3plotsl0} (a) and (b).  In both regions close to the boundary ($x=0$) the zero mode operator is dominated by the single particle component $u^{(1)}(x)$, which  resembles closely the non-interacting wave function and decays exponentially in space with the coherence length $\xi\sim t/\Delta $.  In Region I  we find that the three-particle components made of local clusters of Majorana operators $u^{(3)}_{\rm l}(x)$ are everywhere  smaller than the single particle component $ u^{(1)}(x)$, and follows the same spatial decay, see Figure \ref{fig:u3plotsl0} (a).

Region II shows  larger overall weights in the multi-particle content of the zero modes, see Figure \ref{fig:u3plotsl0} (b). Crucially in this regime we see that local clusters of Majoranas  $u^{(3)}_{\rm l}(x)$ decay much slower than $u^{(1)}(x)$,   in what seems to resemble a power law. The presence of such terms implies that the modes on opposite sides of the system are more strongly coupled. 

{\it Decay rates - Disorder case:}
Disorder has a strikingly different effect on the decay of the different
multi-particle components in the two regimes of parameters. In Region~I [Fig.
\ref{fig:u3plotsl0} (c)], disorder extends  the spatial profile of  $
u^{(1)}(x)$ as well as  $u^{(3)}_{\rm l} (x)$, which both follow  the spatial
decay of the non interacting wave function (dashed black line) close to the
boundary and saturate to a larger  decay length  away from the boundary.
Crucially this transition to a longer decay length does not happen in the clean
wire limit. We see this  as evidence that disorder has driven the system from
Region I to Region II. In Region II we see that disorder reduces the spatial
extent of both the single particle component $u^{(1)}(x)$ and  local clusters
$u^{(3)}_{\rm l} (x)$ from no or power law decay in the clean limit Fig.
\ref{fig:u3plotsl0} (b) to exponential decay  Fig. \ref{fig:u3plotsl0} (d).  
\begin{figure}
\centering
\includegraphics[width=.48\textwidth,height=.2\textwidth]{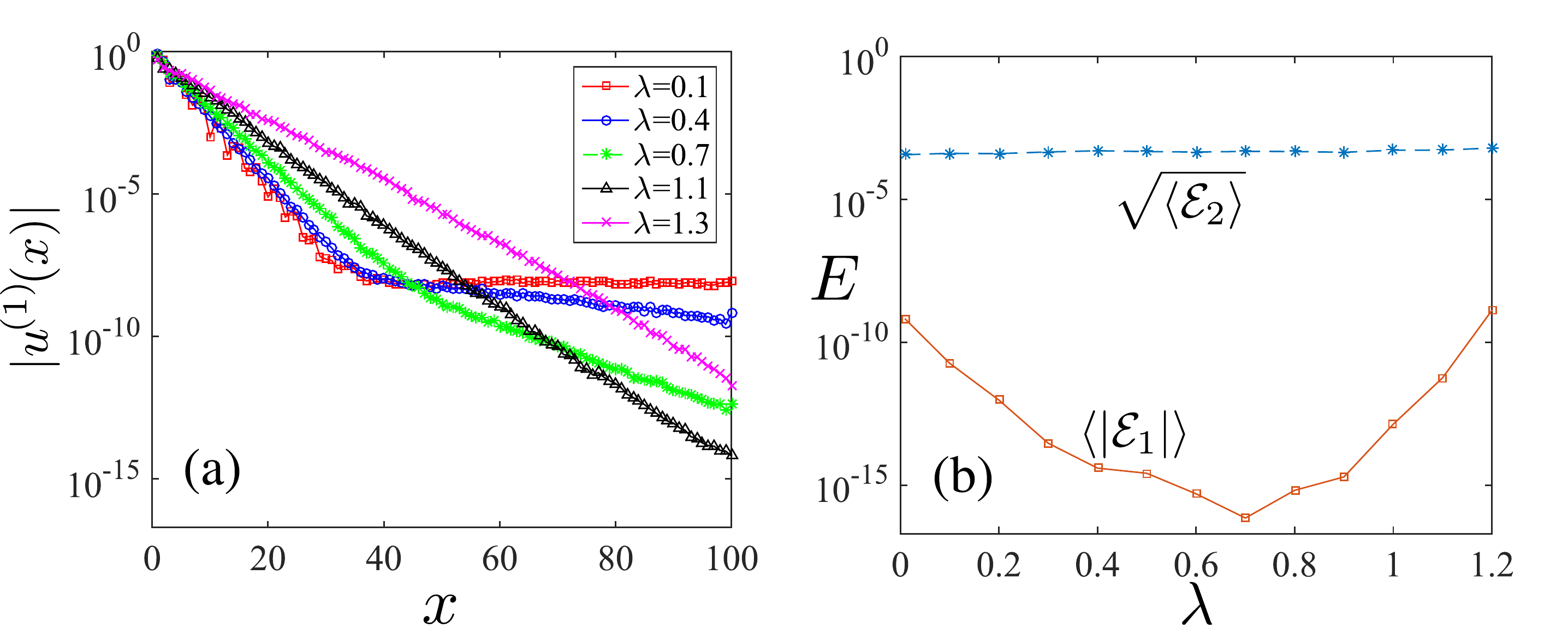}
\label{fig:spEplot}
\caption[]{ (a) The behaviour of the disorder-average single particle components for different amounts of on-site disorder $\lambda$.  The cleaner systems show  an exponential decay  that follows  the non interacting coherence length near the boundary, which revert to a more moderate decay in the bulk.  Disorder typically increases the effective single particle coherence length near the boundary while reducing the residual decay lengths in the bulk. (b) For a long enough wire we can see this behaviour is  correlated with the disorder averaged $\mathcal{E}_1 \sim \langle \delta \rangle$ estimate. In contrast this behaviour is clearly not correlated with the $\langle \mathcal{E}_2\rangle$ estimate, which is  expected to be dominated by non-diagonal disorder $ \chi_2$.  This data is for Region II ($\Delta=0.5t $ and $ \mu=-t$) with $U=0.01t$ and bond dimension $\chi=64$}
\label{fig:sp_and_statistics}
\end{figure}

{\it Energy splitting statistics:}
The previous results outlined how the degree of localization of the approximate zero mode depends on the amount of disorder. Figure \ref{fig:sp_and_statistics} shows the correlation between  these spatial  decay rates and the  mean energy splitting estimate $\mathcal{E}_1 \sim \langle \delta \rangle$ in Region II.  The single particle components show clearly the dual nature of disorder. On a mean field level, disorder extends the effective coherence length. This is  manifested in a moderation of the  exponential decay near the chain edge, which follows the non interacting spatial profile. Conversely, the residual non exponential decay, absent in a non interacting system, is progressively reduced in a disordered medium, see Figure \ref{fig:sp_and_statistics} (a). In a long chain, contributions from these non-exponential tails dominate the mean pairwise energy  splitting $\mathcal{E}_1 \sim \langle \delta \rangle$.  As such the mean energy-splitting is decreased by moderate disorder, see Figure \ref{fig:sp_and_statistics} (b).  As disorder is increased further, the single-particle effect dominates and $\mathcal{E}_1$ increases.  

Figure \ref{fig:sp_and_statistics} (b) also shows the associated $\langle \mathcal{E}_2 \rangle^{1/2}$ estimate. Although this number is expected to be dominated by $\chi_2$ in this regime, it does represent an upper bound on the expected spread of the distribution (see discussion about errors in the Appendix \ref{sect:dmrgsuper}). To address these statistics more directly, in Appendix \ref{sect:ED} we also examine distributions for small systems using exact diagonalisation.  We find that  disorder {\it reduces} the occasional large energy splitting from real decay transitions, but that the standard deviation (about the mean) of pairwise splitting shows only a modest initial decrease with disorder, eventually being overcome due to the {\it increase} in the probability of bands to overlap and/or the single particle effect which dominates near the system edges.  

{\em Discussion of Multi-particle weights:} As a last measure we analyse the  total multi-particle content of the zero mode wave function. 
For this purpose we define the integrated weight in a given $n$-particle sector as  the sum of all $n $ particle terms: $ |N_n|^2  = \int  |u^{(n)} (\vec{x})|^2 d\vec{x}$, which have the property that $\sum_n |N_n|^2 =1$. The total multi-particle content of the zero mode wave function is then given by $1-|N_1|^2$. It has been argued previously that operators with larger weights in these multi-particle sectors decohere more quickly \cite{Goldstein2012}, it could be used as a signature for localization-enhanced topological order, see Ref. \onlinecite{Kells2015b} and Appendix \ref{sect:Majconstruct}.

\begin{figure}
\centering
\includegraphics[width=.48\textwidth,height=0.22\textwidth]{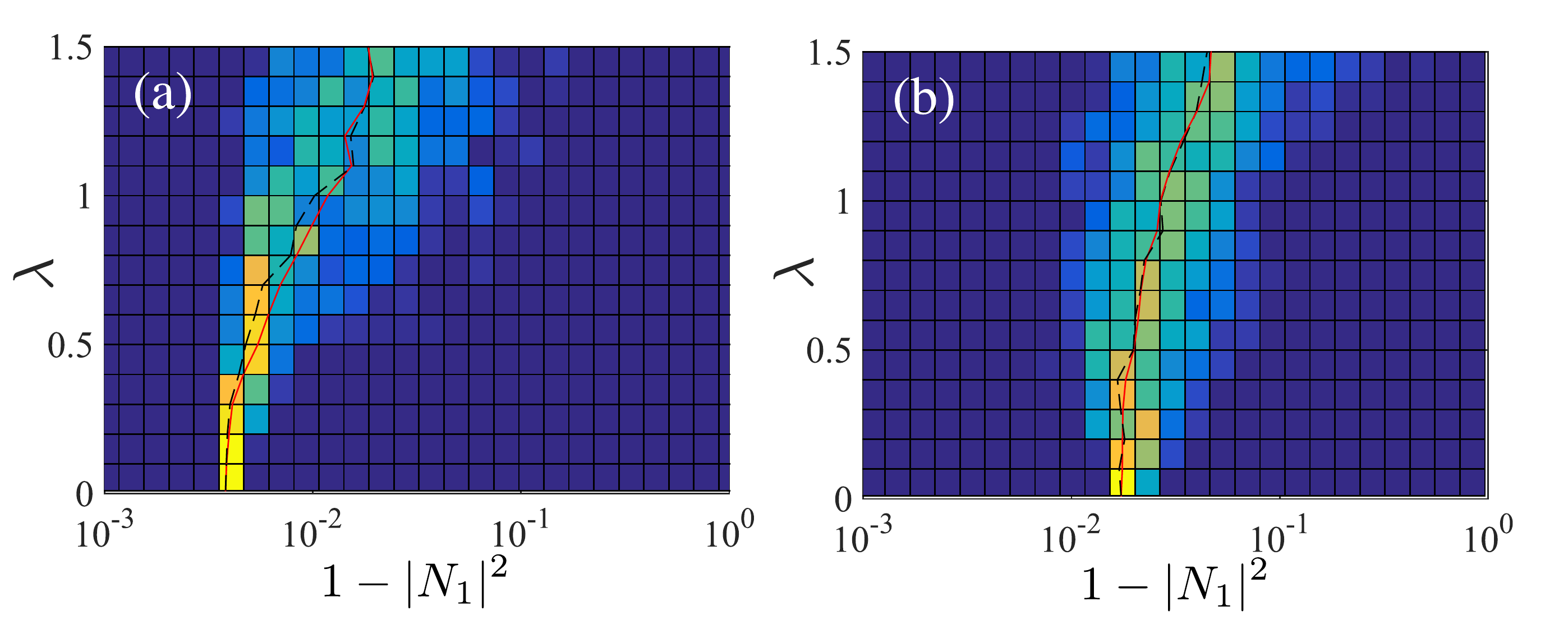}
\caption[]{ Distributions of the multi-particle content  $1-|N_1|^2$ of the approximate steady state as a function of disorder strength $\lambda $, for (a) Region I ($\Delta=0.7t $ and $ \mu=-0.2t$) and (b) Region II ($\Delta=0.5t $ and $ \mu=-t$). The plots were obtained  for a system of length $L=100$,  using an interaction strength $ U=0.1t$ and constant MPS bond dimension of $\chi=128$.   In Region I the $1-|N_1|^2$ probability distribution displays a hard minimum and disorder can only increase the weight supported in the multi-particle sectors. In Region II we see that disorder has a chance to increase or decrease the multi-particle weight.  Mean and median values are shown in red and black dashed lines respectively)  }
\label{fig:1mN1plots}
\end{figure}

Figure \ref{fig:1mN1plots}  shows the distribution of these multi-particle weights as a function of disorder. In Region I we find that disorder  increases the multi-particle weight.  Region II shows a more intricate behavior. Here, while disorder broadens the distribution thus allowing for specific disorder realisations with a smaller multi-particle weight, both the mean (red-line) and median (black-dashed line)  exhibit a monotonic increase. %

{\em Conclusion}
We study numerically the  effect of disorder on the stability of the many body zero mode in a Kitaev chain with local interactions. Our methodology allows us to obtain information about the spatial and multi-particle profile of the zero mode operator, as well as to approximate the statistics of nearly degenerate pairs of states associated with the zero mode, over the entire energy spectrum.  Our analysis shows that the parameter space of a clean system can be divided into regions where relevant interaction-induced decay transitions are  suppressed (Region I)  and  were they are not (Region II).  We find that the effect of disorder on the many body zero mode varies qualitatively between these two regimes. In Region I, disorder has an overall adverse effect: it extends both single particle and multi-particle components further into the bulk while simultaneously increases the likelihood that real decay processes occur.  In Region II disorder has a more intricate effect.  While broadening exponential decay of the single particle components of the zero mode operator, we observe that local-clusters of multi-particles decay more rapidly in a disordered medium. This more rapid decay is reflected in the mean energy splitting between pairs of states of opposite parity which exhibits an overall  {\it reduction} in a disordered medium. 

The qualitative prediction is that these localization effects should also result in a decrease in the width of the energy splitting distribution - resulting in a many-body Majorana operator that is more mode-like. For larger systems we argue that the MPS measure, which could in-principle be used to address the width of the splitting distribution, are in fact dominated by non-zero mode contributions. However, using exact diagonalisation we show that disorder can reduce the size of the occasional large splitting corresponding to real decay transitions. We note however that this effect is counteracted by an increase in the likelihood of these decay transitions occurring and non-interacting single particle effects which tend to dominate near the system's edges.  

We have discussed in some detail how the enhancement effect ties in with the phenomena of many-body-localization. We note that although the underlying mechanisms and the techniques used to study them are similar, there are important distinctions that result in the phenomena being independent of each other.  This can be seen quite clearly in the numerical analysis where we show that measures of MBL cannot resolve the disorder-driven topological phase transition, nor the subtle distinctions between Regions I and II.  We stress however that there is nothing in our work that precludes the coexistence of SPT-order and MBL. 

The MPS/super-operator methodology can be extended to other related models e.g the proximity coupled models or the $\ZZ_n$ parafermionic clock models.  For the class of proximity coupled systems the extension should be possible as these systems also possess a natural non-interacting limit and disorder is known to localise bulk eigenmodes there also.  One possible caveat  is that in proximity coupled systems, the Kitaev-chain arises as an effective low-energy limit, and it is not clear that the  correspondence holds at higher energies. 

For the $\ZZ_n$ models, apart from some special cases (e.g. \cite{Moran2017}) there are no obvious non-interacting limits. There are however a number of works that point to exactly-solvable/free-fermion ground-states  \cite{Iemini2017, Meichanetzidis2017}  Moreover, there are clear indications that special points of clean $\ZZ_n$ models, where $n$ is prime, can contain strong-zero-modes.  One particularly strong candidate for this is the so called $\pi/6$ point in the $\ZZ_3$ system \cite{Jermyn2014,Moran2017, Else2017}.  These special points are natural analogs of Region I and so we expect that these models should share many of the same features with the interacting Kitaev chain.

{\em Acknowledgments}
The authors acknowledge  Yevgeny Bar Lev, Piet Brouwer, Dimitri Gutman, Arbel Haim, Falko Pienkta, Domenico Pellegrino,  Maria-Theresa Rieder, Alessandro Romito and Joost Slingerland for fruitful discussions. G.K. acknowledges support from Science Foundation Ireland under its Career Development Award Programme 2015, (Grant No. 15/CDA/3240). N.M. acknowledges financial support from Science Foundation Ireland through Principal Investigator Award (Grant No. 12/IA/1697). D. M. acknowledges support from the Israel Science Foundation (Grant No. 737/14) and from the People Programme (Marie Curie Actions) of the European Union's Seventh Framework Programme (FP7/2007-2013) under REA grant agreement No. 631064.  We also wish to acknowledge the SFI/HEA Irish Centre for High-End Computing (ICHEC) for the provision of computational facilities and support.

\appendix
\section{Perturbative zero-modes for finite wires}
\label{sect:perturbzero}
The argument given in Ref. \onlinecite{Kells2015a} shows that the existence of the non-interacting Majorana mode also places restrictions on the form of all couplings between eigenstates of the non-interacting system. The argument relies on the fact the perturbative terms due to local parity-preserving operators will be identical in each sector, provided the states involved have the same zero-mode occupation.  As a result, the degenerate perturbation expansions of the bands themselves will look the same in each sector, to an order of perturbation theory that scales with the length of the system ( see also Ref. \onlinecite{Moran2017} ).  However,  this argument does not account for situations where the bands with different fermion number start to hybridize. In this case interaction-induced transitions between bands with different occupation of the non-interacting zero mode will factor into left $\pm$ right  (for even and odd sectors)  \cite{Kells2015a} and therefore when these bands start to intersect we see some even-odd sectoral dependency at avoided level crossings. 

This basic argument allows us to place some more simple limits on the degree to which degeneracy  is protected in the wire.  The single particle spectrum of the non-interacting system is given as  $ \epsilon_k= ( ( -\mu -2t \cos k) ^2 +(2 |\Delta| \sin k )^2)^{1/2} $.  For $\mu<0$, such that we move the chemical potential towards the bottom of the band, the maximum bulk-excitation energy is $\epsilon_{\text{max}} = 2t -\mu$, and the minimum is 
\be
  \epsilon_{\text{min}}=
    \begin{cases}
          2t+\mu, & \text{if } \frac{t}{2} (u+2t) < |\Delta|^2 \\
      |\Delta|  \sqrt{4-\frac{\mu^2}{(t^2 -\Delta^2)}}, & \text{otherwise}
    \end{cases}
\ee
The condition that there are no overlaps between bands that differ by one bulk fermion-excitation is that $ \epsilon_{\text{max}} <2 \epsilon_{\text{min}} $ and we arrive at the inequality which, in the main text, defines our Region I:
\bea
\label{eq:R1def}
 && -\mu  < \frac{2}{3} t, \quad \quad \quad \quad \quad  \quad \text{if } \frac{t}{2} (u+2t) < |\Delta|^2\\
 && -\mu  < 2 ( |\Delta| \sqrt{4- \frac{\mu^2} {t^2 -|\Delta|^2}} -t) ,  \text{otherwise}  \non
\eea

We can also estimate when this spread becomes large enough to close the gap $(2t)$ between the $N^{th}$ and $N-1^{th}$ bands.  Near the flat band limit ($t=|\Delta|$, $\mu=0$), and with $\mu<0$, the maximum of the band occurs at $\epsilon_{\text{max}} = 2t - \mu $ at  $ k=\pm \pi $ and the minimum occurs roughly at $2 |\Delta| \sin k_F$ where $k_F = \cos^{-1} \mu/2t  \approx \pi/2$ and therefore  $\epsilon_{\text{min}} \approx 2 |\Delta| $.  The spread in the single particle spectrum is therefore  $|\mu| + \kappa$, with $\kappa \equiv 2 (t-|\Delta|)$. Assuming we are in a large enough system such that the $N^{th}$ largest and smallest single particle eigenvalues are almost the same we can write the requirement that the bands don't overlap : $N \epsilon_{\text{min}} - (N-1) \epsilon_{\text{max}}  > 0$, which after rearranging becomes
\begin{equation}
\frac{\kappa+|\mu|}{4t+|\mu|} < \frac{1}{N}.
\end{equation}
The condition is restrictive. Close to the middle of the spectrum this occurs at  at progressively small $\kappa $ and $\mu $. A caveat to this however is that the splitting that  occurs between the bands $N$ and $N-1$ (recall that one of these states has an occupied zero mode which we are not counting) comes about because of non-zero matrix elements between states that differ by $\sim 2 N$ fermions.  As such, the interaction-induced transition that couples these states would therefore result in an even-odd splitting of the order $U^{N/2} $ occurring at this interaction induced avoided level crossing. 

Moreover, for a system of length $L$ as we vary $\mu$ or $\kappa$ away from the special point, the first crossing occurs between the $N=L/2$ and say $=N-1=L/2-1$ bands. However as the avoided level crossing here must be proportional to $U^{L/4}$ the even-odd splitting will strictly speaking still be exponential in a parameter that is a sizeable fraction of the system length. The question of whether there is a strong zero mode when both $\mu$ and $\kappa$ are non-zero is therefore a complicated one, and the answer has to be qualified  based on where exactly one is in the parameter space.
  
For a finite wire,  we see that there is a finite region of parameter space for non-zero $\mu$ and $\kappa$ such that there is a strong zero mode. Nonetheless, this region  diminishes  as one approaches the thermodynamic limit. On this point, we note that it is always possible to make the zero mode exact with some small local tweak in parameters near one of the wire ends and thus for  many purposes in what follows it is useful to proceed as if there is an exact zero mode and to explore the consequences that this must have for it's multi-particle content. 

\section{DMRG for superoperators}
\label{sect:dmrgsuper}
Our algorithm attempts to construct multinomials of position space Majorana operators:
\begin{align}
%\label{eq:Majspl}
\gamma_L(U) = \sum_i^{2L} u^{(1)}_L(i) \gamma_i + \sum_{ijk}^{2L} u^{(3)}_L(i,j,k) \gamma_{i} \gamma_{j} \gamma_{k} + ... \\
\gamma^R(U) = \sum_i^{2L} u^{(1)}_R(i) \gamma_i + \sum_{ijk}^{2L} u^{(3)}_R(i,j,k) \gamma_{i}  \gamma_{j} \gamma_{k} + ... \non
\end{align}
 that almost commute with the interacting Kitaev wire Hamiltonian.  These modes are normalised such that if we define weights $N^{n} =\int |u^{(n)}(\vec{x}|^2 dx$ then $\sum N^{n} =1 $ .  In the non-interacting system, the expansion coefficients $u^{(n)}=0$ for all $n>1$, and therefore $N^{(1)}=1$.   In keeping with the idea of the mode as a dressed quasi-particle,  we expect that the single particle weight $N^{(1)}$ dominates the other multi-particle weights also in the case of non-zero interaction strength $U$.  
 
In Ref. \onlinecite{Kells2015b} it was demonstrated how one can approximate such a mode using a real space  approach that selectively sampled the multi-particle components that were close, in configuration space, to single particle operators.  The technique works by creating a matrix representation for the super-operator $\mathcal{H} = [ H, \bullet]$ and finding approximate steady states of the form \eqref{eq:Majspl} by variationally approaching a single-particle dominated null vector of $\mathcal{H}^2$.  

One difficulty with this method is that the Hilbert space dimension of null-vectors of $\mathcal{H}$  grows as $2^L$ and moreover is itself embedded in a continuum.  However,  it is possible to argue that within this continuous band of excitation energies there are only two approximate steady-states with the form  \eqref{eq:Majspl} that are dominated by the single-particle elements.  Moreover, by continuity  it is straightforward to argue that the single particle components of the operators should have a similar structure to their non-interacting counterparts.   

The variational step searches for null vectors of $\mathcal{H}^2$ using a Lanczos algorithm. The initial states for the procedure are the non-interacting Majorana's on both ends of the wires.  Working with $\mathcal{H}^2$ is needed to ensure our eigenvalue approximation is bounded from below, and has the additional advantage that this operator preserves sub-lattice symmetry.   Thin restarting is needed to ensure that on each iteration of the algorithm the updated state resembles the input state.  

The algorithm that we use in this paper can be seen as a hybrid of the aforementioned real-space sampling approach and methods that seek to use DMRG approaches to approximate the null-vectors of $\mathcal{H}$ or more generally the Limbladlian \cite{Mascarenhas2015,Cui2015}.  The key difference with the real-space sampling approach is that the operator $\mathcal{H} = [ H, \bullet]$ is now represented as a Matrix-Product-Operator.  From this we contract indices of the MPO to obtain an MPO for $\mathcal{H}^2$ and then search for its null-vectors using a modified DMRG sweeping procedure. 

To ensure that algorithm converges to the single-particle dominated modes we found it necessary to again employ Lancsoz thin-restarting, this time at each optimisation step in the sweep along the wire/chain.   In terms of overall efficiency we note that orders of magnitude improvement can be obtained by also implementing a controlled compression of the MPO $\mathcal{H}^2$.  

\subsection{Discussion of numerical errors in the MPS variational technique}

In the main text we argued that the MPO/MPS representation of the zero-mode could be written in the eigenbasis of the Hamiltonian as:
\begin{align}
\Gamma_{L} &=  \sum_{n} (1-\alpha^{L}_n) \left[ |n_0\rangle \langle n_1| + |n_1\rangle \langle n_0| \right]\\ 
\non &+  \sum_{n \ne m} \beta^L_{nm}  \left[ |n_0\rangle \langle m_1|+ |m_1\rangle \langle n_0| \right]\\
\Gamma_{R} &=-i \sum_{n} (1-\alpha^{R}_n) \left[ |n_0\rangle \langle n_1| - |n_1\rangle \langle n_0| \right]\\ 
\non & -i \sum_{n \ne m} \beta^R_{nm}  \left[ |n_0\rangle \langle m_1| - |m_1\rangle \langle n_0| \right]
\end{align}
where  the $\alpha$ and $\beta$-terms represent diagonal/off-diagonal errors respectively.  Moreover we showed that the estimates for the energy level statistics are calculated using the trace formula:
\begin{align}
\label{eq:ee1}
& \mathcal{E}_1 =i \Tr( \Gamma_L \times [H, \Gamma_R] )/2^L =  \langle \delta \rangle + \chi_1 \\
\label{eq:ee2}
& \mathcal{E}_2 =\Tr( \Gamma_L\times[ H, [H, \Gamma_L]] )/2^L = \langle \delta^2 \rangle   +\chi_2
\end{align}
where
\begin{align} 
\label{eq:chi1}
\chi_1&= - \frac{1}{2^L}  \sum_n \delta_n \left[\alpha^{L}_n +\alpha^R_n  - \alpha^{L}_n \alpha^{R}_n \right] \\ 
\non &+  \frac{1}{2^L} \sum_{n\ne m} \beta^L_{nm} \beta^R_{nm} [E_n-E_m+\frac{\delta_n}{2}+\frac{\delta_m}{2}] 
\end{align}
and
\begin{align}
\label{eq:chi2}
\chi_2&= - \frac{1}{2^L}  \sum_n  \delta_n^2 ( 2 \alpha^L_n- (\alpha^L_n)^{2}) \\ 
\non&+  \frac{1}{2^L} \sum_{n\ne m}( \beta^L_{nm})^{2} [E_n-E_m+\frac{\delta_n}{2}+\frac{\delta_m}{2}]^2. 
\end{align}
Here, the error $\chi_1$ is an average over small contributions of random sign. We therefore expect it to be negligible. 
In contrast $\chi_2 $  can dominate the second moment  $\mathcal{E}^2$. 

To check this conjecture,  we perform the following simple test. At fixed bond dimension we compare the asymptotic value of the $\mathcal{E}_2$(calculated using the $\Gamma_L$ operator) with the  asymptotic value of a almost identical set up, given by Eq. \eqref{H:Kitaev} and \eqref{H:Int} where the coupling terms on the very right hand side of the system are changed to ensure we have a perfectly decoupled $\Gamma_R = \gamma_{2N}$ Majorana (e.g we set $\mu_N=0$, $\Delta_{N-1}= t \arg(\Delta)$ and $U_{N-1}=0$ ). In the modified setup,  as a result of the perfectly decoupled $\Gamma_R$ Majorana, the  true many body spectrum is exactly two fold degenerate, which corresponds to Eq. \eqref{H:many_body}, with  $\delta_n=0 $ for all $n$. In this setup,  the first and second moments are determined solely by the off diagonal errors in $ \chi_1 $ and $\chi_2 $:
\begin{align} 
 \mathcal{E}_1= \chi_1&=  \frac{1}{2^L} \sum_{n\ne m} \beta^L_{nm} \beta^R_{nm} [E_n-E_m] 
\end{align}
and
\begin{align}
 \mathcal{E}_2=\chi_2&=   \frac{1}{2^L} \sum_{n\ne m}( \beta^L_{nm})^{2} [E_n-E_m]^2. 
\end{align} 
 
Moreover, in the modified setup, the right Majorana can be determined exactly, and  $ \alpha^R=\beta^R=0$. Consequently,  the resulting first moment $\mathcal{E}$  will be identically zero, regardless of the numerically calculated  $\Gamma_L$.  
 
In  contrast the estimate for $\mathcal{E}_2$  are affected by errors of the calculated $\Gamma_L$ only and do not necessarily vanish. We find that the estimate for the second moment $\mathcal{E}_2$ in the modified setup with an exact degeneracy and in the original setup given by Eq. \eqref{H:Kitaev} and \eqref{H:Int} to be comparable. This supports our conjecture that the second moment calculations are dominated by off diagonal errors. 
 
\begin{figure}
\centering
\includegraphics[width=.48\textwidth,height=0.24\textwidth]{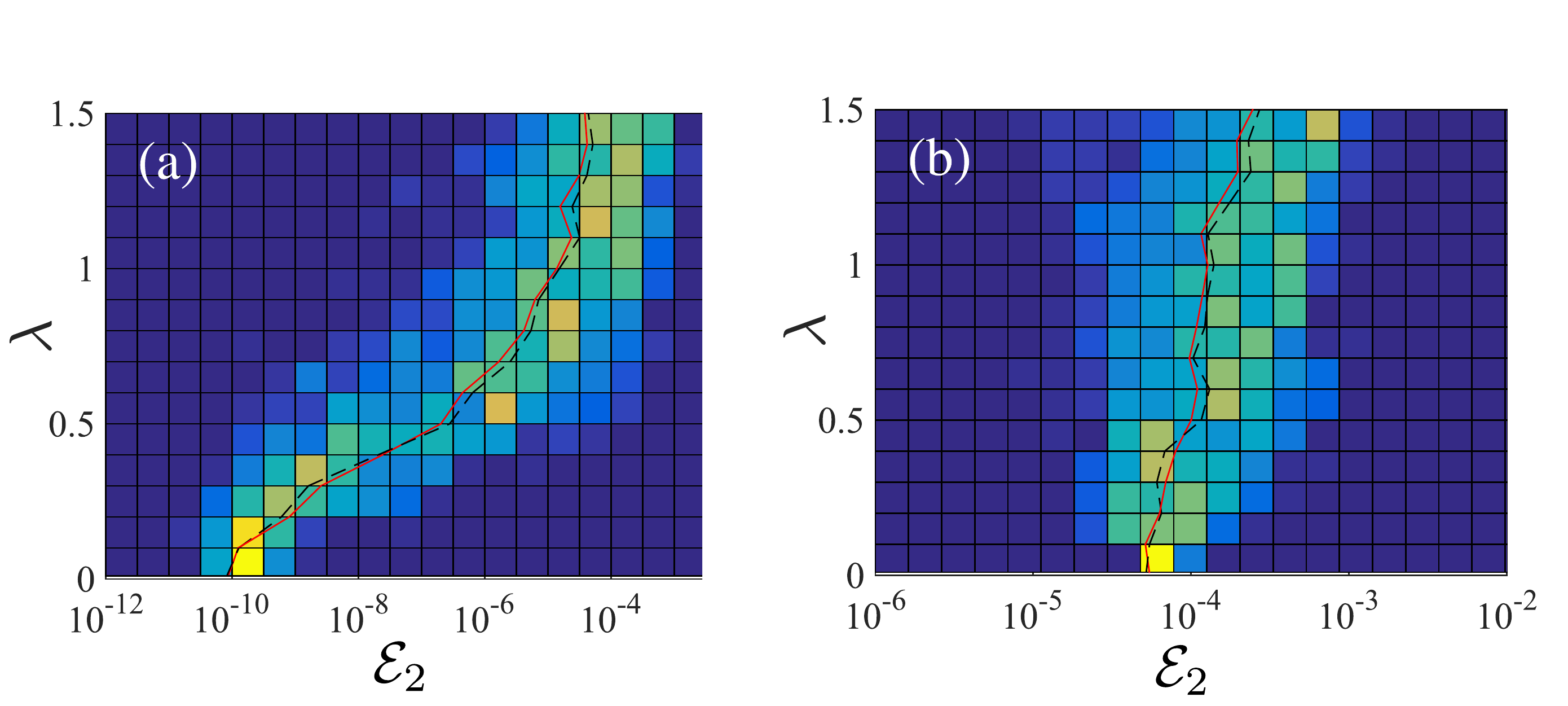}
\caption[]{ Distributions of  $\mathcal{E}_2$ as a function of disorder strength $\lambda $, for (a) Region I ($\Delta=0.7t $ and $ \mu=-0.2t$) and (b) Region II ($\Delta=0.5t $ and $ \mu=-t$). The plots were obtained  for a system of length $L=100$,  using an interaction strength $ U=0.1t$ and constant MPS bond dimension of $\chi=64$.  One hundred disorder realisations are used for each value of $\lambda$. In Region I the dominant effect of disorder is to push the system towards values closer to those observed in Region II. In Region II we see that disorder has a chance to increase or decrease the $\mathcal{E}_2$. We observe that the average and median values (red and black dashed lines resp.) tend to increase as we increase the amount of disorder.
}
\label{fig:e2plots}
\end{figure}

Figure \ref{fig:e2plots} shows the distribution of $\mathcal{E}_2 $  as a function of disorder in (a) Region I and (b) Region II.
A similar calculation in the modified setup gives comparable results which indicates that the value of $\mathcal{E}_2 $  is dominated by off diagonal error $\chi_2 $. Moreover, the data taken in the modified setup show a similar trend with disorder. This suggest that $\mathcal{E}_2 $ and consequently  $ \chi_2$ generally reflect the extent of mixing between bands. In this respect we see for example that in Region I disorder substantially increases the $\mathcal{E}_2 $  estimate, indicating that it drives the system into the regime where real transitions can occur. In Region II the $\mathcal{E}_2$ estimate shows a moderate increase with disorder, in accordance with the expectation that it is determined by the degree of mixing between bands.

\section{Measures of localization enhanced topological order in Exact-Diagonalisation calculations}
\label{sect:ED}

\begin{figure}[!htb]
 \centering
\includegraphics[width=0.45\textwidth,height=0.18\textwidth]{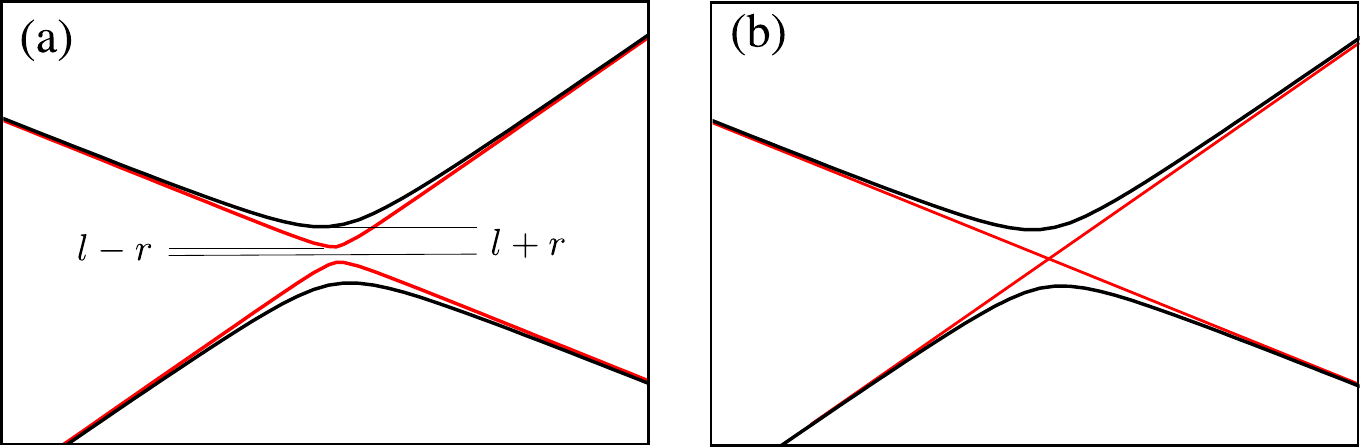}
\caption{ At crossings between states with different (approximate) occupations of the zero-mode, the interaction dependent splitting coefficient factors in to a left and right contribution (see e.g. \cite{Kells2015a}).  In the presence of disorder, the average splitting coefficients $l$ and $r$  should decay exponentially with the system length $L$.  (b) In a clean system with perfect reflection symmetry about the centre of the wire one of the sectors,  depending on the total fermionic and spatial parity, will always display a symmetry protected crossing.   }
 \label{fig:AVC}
 \end{figure}

Interaction induced decay transitions, which can change the occupancy of the zero mode  while exciting Bogoliubov quasi-particles, occur when bands of different fermion occupation number cross. The key prediction of localization enhanced topological order, is that the resulting mismatch in even-odd energy levels at the avoided crossings (see Figure \ref{fig:AVC}) will become smaller as the amount of disorder is increased.  

In exact diagonalisation this effect is quite difficult to discern in the overall energy level splitting statistics. This is  because, in the parameter space accessible to exact diagonalisation (ED) (large $\Delta$), disorder increases (on average) both the overall splitting of pairs as well as the probability that an anomalous splitting can occur.  The effect on the average splitting can be  understood on a single particle level as  resulting from the increases in the effective coherence length in a disordered medium.  The increase in the number of anomalous splittings comes about because disorder will also broaden the bands and hence increase the chances that bands with different fermion number overlap. Hence, while disorder reduces the value of anomalous pair splitting, it  increases their number. 

\begin{figure}
\centering
\includegraphics[width=0.42\textwidth,height=0.32\textwidth]{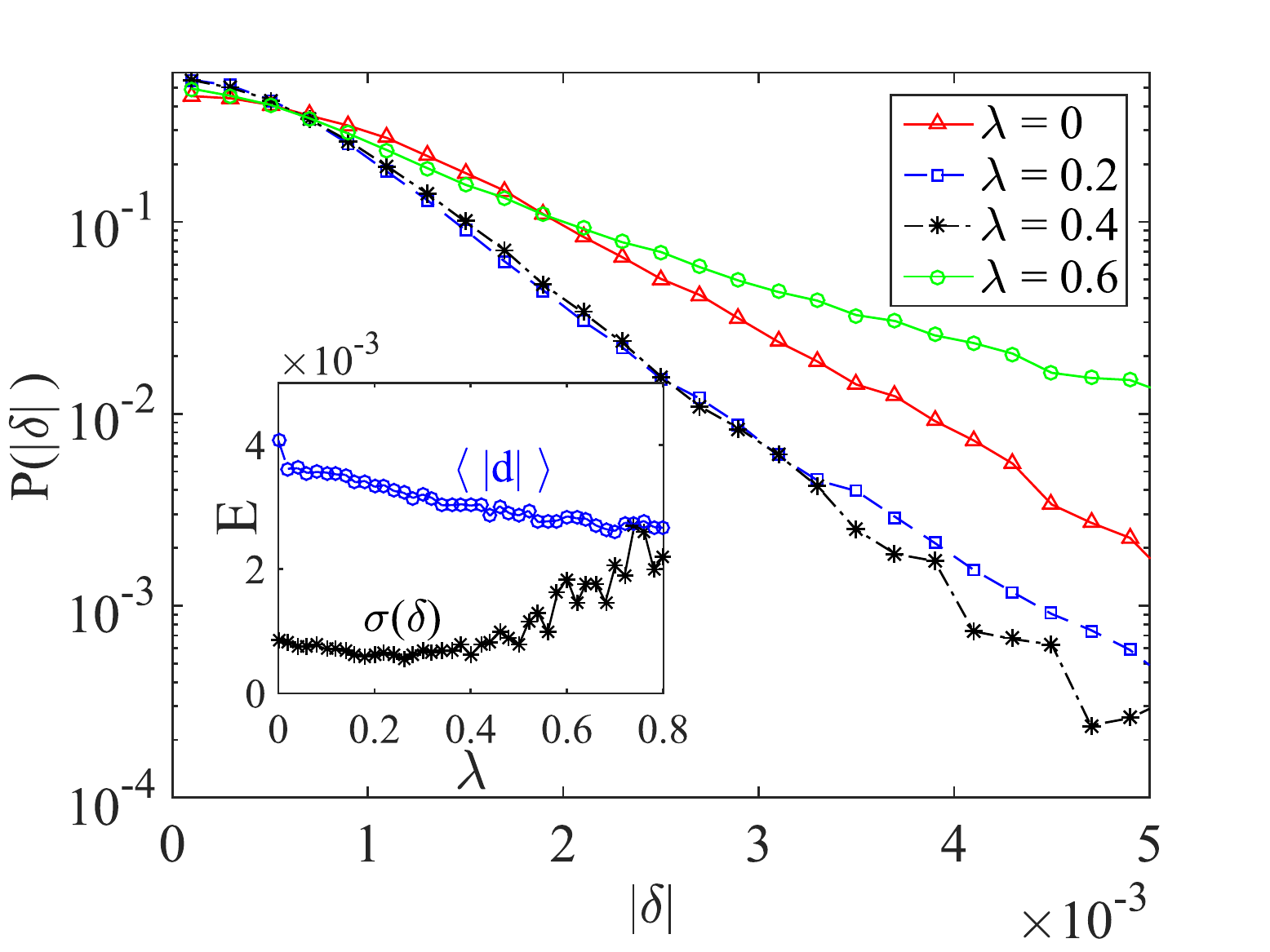}
\caption{ For small system sizes the disorder averaged distribution $P(\delta)$ can become less broad with a modest amount of disorder. This is because, even for very small systems sizes ($L=13$) the matrix elements responsible for real decay processes are  reduced as a function of disorder parameter $\lambda$.  The effect quickly disappears as one increases disorder because this (on-average) drives the system further into Region II and also increase the single-particle coherence length of the Majorana end-states (an effect which is more relevant for short systems).  The inset shows the standard deviation $\sigma(\delta)$, calculated using the full spectrum, and the $\langle | d| \rangle$ is the mean of {\em all} interacting matrix elements (in the non-interacting basis) that differ by occupation of the zero-mode .  The data was generated in Region II, with $\mu=-1, \Delta=0.5t$ and $U=0.1t$.  }
 \label{fig:hL13}
 \end{figure}

For small system size, the latter explains qualitatively why one should not necessarily observe a reduction in the global statistical quantities such as $\text{Var}(\delta)$ even though the responsible matrix elements should be reduced by disorder (see Figure \ref{fig:hL13})  for the corresponding calculation for $L=13$).  The ED calculations, however,  allow us to calculate the entire probability distribution of energy levels. Here the reduction in decay like transitions,  becomes apparent in the shape of the distributions at higher-pair splitting.  

\section{Formal Majorana construction and the connection between the energy splitting and multi-particle content}
\label{sect:Majconstruct}
In terms of the eigenstates of the system we can write
\begin{align}
\label{eq:MBM}
\gamma_L(U) &= \phantom{i} \sum \ket{n_0} \bra{n_1} + \ket{n_1} \bra{n_0}  \\
\gamma_R (U) &= i \sum \ket{n_0} \bra{n_1} - \ket{n_1} \bra{n_0}  .
\end{align}
where the sub-script the approximate occupation of the zero-mode.  This method of constructing the modes is formally identical to the method of {\em l}-bit construction in the MBL literature, see for example Ref. \onlinecite{Huse2014}, and requires one to be able to identify pairs of states  $\ket{n_0}$ and $\ket{n_1}$ ,  and then to fix the relative phases.  One way to make this identification in principle is to use the energy of the states as an identifier and match states according to where they occur in the energy spectrum.  Another identification method is to examine  how well two states are mapped to each other by the non-interacting modes,which can be calculated exactly:
\begin{equation}
O_{L/R}=\bra{n_\text{even}} \gamma_{L/R}(0) \ket{m_\text{odd}} 
\end{equation}
This latter method also allows one to determine the correct relative phase.

\begin{figure}
\includegraphics[width=0.4\textwidth,height=0.18\textwidth]{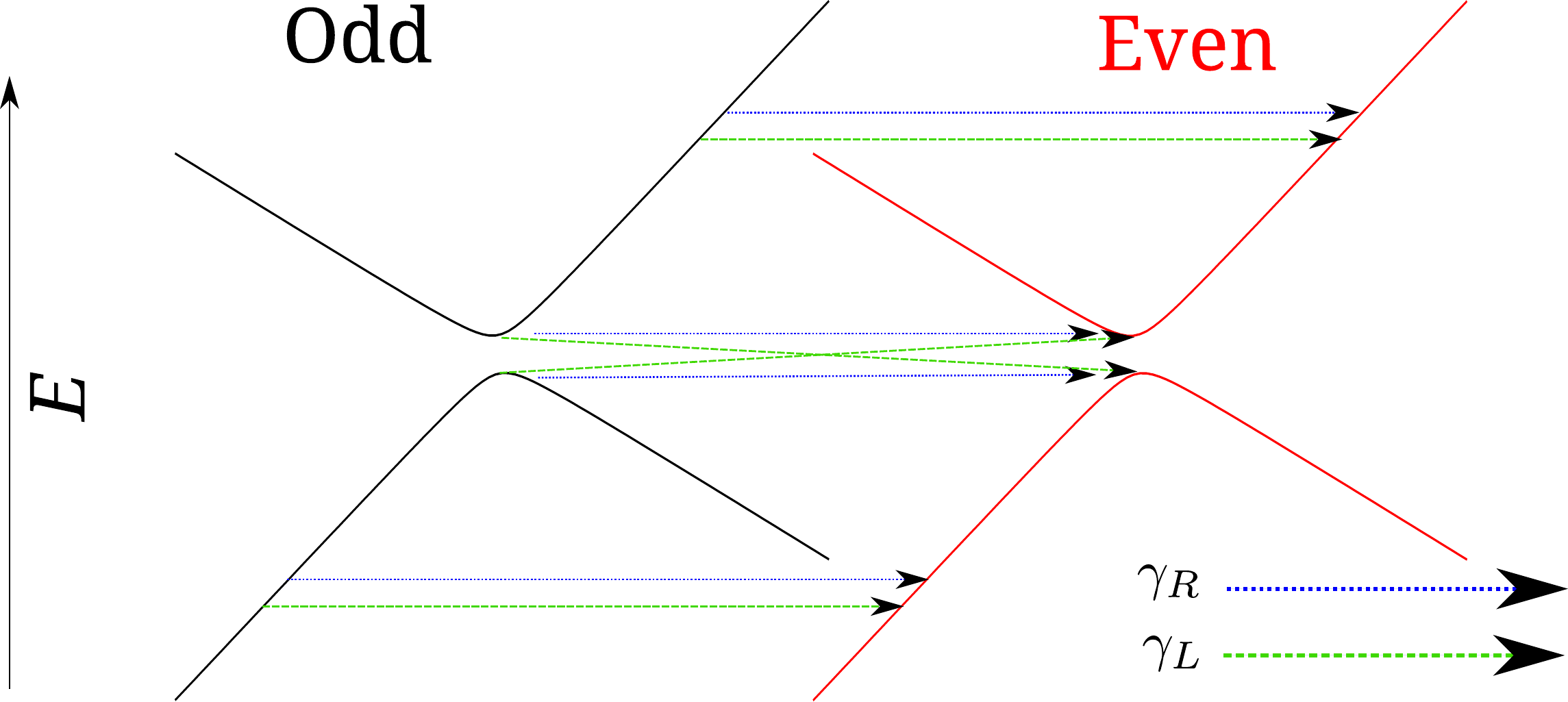}
\caption{ In order to construct the many-body Majorana zero mode one needs to first identify which states in each sector to pair up. Away from the avoided crossings this identification can be achieved either by examining the energy or by using the non-interacting Majorana modes $\gamma_{L/R}(0)$. At the crossing however, one of the non-interacting modes (in this figure we have chosen $\gamma_L(0)$) will always identify states at the other side of the crossing. This sets up the relationship between multi-particle content and energy. }
 \label{fig:AVCidentify}
 \end{figure}

In the case of well separated fermionic bands both identification criteria ($E$ and $O_{L/R}$) are in agreement.   However, band crossings may introduce  
an ambiguity between these methods of identification .  This has implications for the zero-mode's residual energy and its multi-particle content. 

To see this, consider what happens at an avoided crossings between states from the same parity sector that differ in their occupation of the zero mode. Working in the basis  $\ket{n}_{0/1}$, $\ket{n+1}_{1/0}$, we can understand the crossing point using the following parametrization of the matrix elements between the relevant states
\begin{equation}
H_c=E_c +\left[ \begin{array}{cc}   a s  & l \pm r  \\ l \pm r  & b s  \end{array} \right] 
\label{eqn:sigma}
\end{equation}
where $s$ is related to the parameters of the non-interacting Hamiltonian ($\mu$, $\Delta$, $t$), and $a$ and $b$ are the slopes of the energy levels at the crossing.  Here the off-diagonal elements $d_{e,o}= l \pm r$ are the interaction-induced coupling coefficients in the even and odd parity sectors, respectively, which are generally different. The partition into left (l) and right (r)  components comes about  because the states in question differ in the occupation of the fermionic zero mode $\beta_0^\dagger\beta_0$ and thus we need to operate with either $\beta_{0}=\gamma_L+i\gamma_R$ or $\beta^\dagger_{0}=\gamma_L-i\gamma_R$  to connect them.

The question we now ask is which states are identified as pairs by the non-interacting Majoranas $\gamma_L(0)$ and $\gamma_R(0)$?  Away from the crossings the relevant eigen-subspace is $\ket{\psi_e}= \{ \ket{n}_0, \ket{n+1}_1 \}$ and $\ket{\psi_o}= \{ \ket{n}_1, \ket{n+1}_0 \}$ where $e/o$ denotes the even/odd sector and $0/1$ denote the approximate occupation of the zero mode. In this basis we have
\begin{equation}
\non \bra{\psi_e} \gamma_L (0)\ket{\psi_o}  \sim I \quad  \bra{\psi_e}\gamma_R(0) \ket{\psi_o}  \sim \sigma^z.
\end{equation}
At  the crossing, the interaction lifts the degeneracy and the modified  eigenbasis is  rotated to symmetric and antisymmetric combinations: $\ket{\tilde{\psi}_e}= \{ \ket{n}_0 \pm \ket{n+1}_1 \}$ and $\ket{\tilde{\psi}_o}= \{ \ket{n}_1 \pm \ket{n+1}_0 \}$.  Therefore in this scenario one of the non-interacting Majoranas will always identify with a state at the other side of the crossing:  For the particular example when 
$l \pm r>0$ we get
\begin{equation}
\non  \bra{\tilde{\psi}_e} \gamma_L (0)\ket{\tilde{\psi}_o} \sim I \quad  \bra{\tilde{\psi}_e} \gamma_R(0) \ket{\tilde{\psi}_o} \sim \sigma_x,
\end{equation}
implying that the non-interacting left Majorana connects states in opposite parity sectors, with an energy splitting of the same sign $ \pm\Delta_{e}\xrightarrow{\gamma_L} \pm\Delta_{o} $ while  the non-interacting right Majorana identifies  states with energy splitting of an opposite sign $ \pm\Delta_{e}\xrightarrow{\gamma_R} \mp\Delta_{o} $.  (In cases when $l \pm r<0$ results in a similar scenarios where  it is $\gamma_L(0)$ that identifies states on opposite side of the crossing). 

More insight can be gained by considering the scenario where one artificially forces one side  of the system (say the right) to be non-interacting. This ensures that  the single-particle  $\gamma_R(0)$ Majorana is an exact zero-mode. Moreover, in this scenario all the interaction-induced coupling coefficient $r$ at the r.h.s vanish identically, and the splitting in both sectors is identical. By construction, $\gamma_R(0)$ will identify states at the same energy, even in the case where there is an avoided crossing.  However, at the crossing point, the non-interacting Majorana at the left hand side will identify  states with an energy mismatch, see again Figure \ref{fig:AVCidentify}.

This highlights the ambiguity between constructing a near-zero mode operator which pairs up states of opposite parity with a minimal energy mismatch, or near zero mode which operates similarly to  its non interacting counter part. As an example, in the modified setup described above, we may construct an operator $\gamma_L(U)$  using states $\ket{n}_e$ and $ \ket{n}_o$ which are identified by the non interacting $\gamma_L(0)$. The resulting operator would connect states with an  energy mismatch - in other words it is not a zero mode. On the other hand,  constructing an operator out of states with  identical energies would result in  a zero-mode by definition, but the states chosen to appear together in the outer product are very different from the ones that are suggested by the single-particle $\gamma_L(0)$. As such, we can construct a zero-mode but with the price that the operator does not resemble the non-interacting mode within this subspace.


\begin{thebibliography}{}
\bibitem{clusters} There is no particular reason for using this choice of local/non-local clusters. Other choices yield the same general picture.
\end{thebibliography}
\end{document}